\documentclass[journal]{IEEEtran}

\ifCLASSINFOpdf
\else
   \usepackage[dvips]{graphicx}
\fi
\usepackage{url}

\usepackage{graphicx}
\usepackage{cite}
\usepackage{amsmath}
\usepackage{amssymb}
\usepackage{bm}
\usepackage{array}
\usepackage{xcolor}

\usepackage{ulem}

\begin{document}

\title{Self-boosted Weight-Constrained FxLMS: A Robustness Distributed Active Noise Control Algorithm without Internode Communication}

\author{Junwei Ji,~\IEEEmembership{Student Member,~IEEE,}
        Dongyuan Shi,~\IEEEmembership{Senior Member,~IEEE,}
        Zhengding Luo,~\IEEEmembership{Student Member,~IEEE,}
        Boxiang Wang,~\IEEEmembership{Student Member,~IEEE,}
        and Woon-Seng Gan,~\IEEEmembership{Senior Member,~IEEE}
\thanks{Junwei Ji, Zhengding Luo, Boxiang Wang and Woon-Seng Gan are with the School of Electrical and Electronic Engineering, Nanyang Technological University, Singapore 639798, Singapore (e-mail: JUNWEI002@e.ntu.edu.sg).}

 \thanks{Dongyuan Shi is with the Center of Intelligent Acoustics and Immersive Communications, Northwestern Polytechnical University, Xi'an, China 710071 (e-mail: dongyuan.shi@nwpu.edu.cn).}

\thanks{This work was supported by the Ministry of Education, Singapore, through Academic Research Fund Tier 2 under Grant MOE-T2EP20221-0014 and Grant MOE-T2EP50122-0018.}
 
 }

\maketitle

\begin{abstract}

Compared to the conventional centralized multichannel active noise control (MCANC) algorithm, which requires substantial computational resources, decentralized approaches exhibit higher computational efficiency but typically result in inferior noise reduction performance. To enhance performance, distributed ANC methods have been introduced, enabling information exchange among ANC nodes; however, the resulting communication latency often compromises system stability. To overcome these limitations, we propose a self-boosted weight-constrained filtered-reference least mean square (SB-WCFxLMS) algorithm for the distributed MCANC system without internode communication. The WCFxLMS algorithm is specifically designed to mitigate divergence issues caused by the internode cross-talk effect. The self-boosted strategy lets each ANC node independently adapt its constraint parameters based on its local noise reduction performance, thus ensuring effective noise cancellation without the need for inter-node communication. With the assistance of this mechanism, this approach significantly reduces both computational complexity and communication overhead. Numerical simulations employing real acoustic paths and compressor noise validate the effectiveness and robustness of the proposed system. The results demonstrate that our proposed method achieves satisfactory noise cancellation performance with minimal resource requirements.
\end{abstract}


\IEEEpeerreviewmaketitle

\section{Introduction}

\IEEEPARstart{A}{ctive} noise control (ANC) is an advanced technique designed to effectively attenuate low-frequency noise that is challenging to manage using traditional passive methods \cite{Kuo1999ANC,Elliott2001SPAC,Kajikaea2012ANC}. By generating an anti-noise signal with the same amplitude but opposite phase to the unwanted noise \cite{lueg1936process}, ANC can mitigate noise-induced physical and psychological health issues \cite{peris2020noise}. Recently, several ANC algorithms have been developed to meet various application situations \cite{Lam2021Ten}, such as filtered reference least mean square (FxLMS) \cite{Morgan1980FXLMS,Bjarnason1992MFXLMS} for adaption, output-constrained methods for avoiding saturation \cite{Shi2019TwoGDFXLMS,Shi2021OLFXLMS,Ji2025MOV}, wireless technique for enhancing signal quality \cite{shi2020simultaneous,luo2022ec,Shen2022MANCwireless}, virtual sensing for limited sensors placement \cite{liu2024relative,sun2022spatial,wang2025transferable}, and deep learning method for slow adaption and non-linearity \cite{zhang2021deep,Luo2023GFANC,xie2024cognitive}.

In recent years, multichannel active noise control (MCANC) systems have drawn great attention since they can achieve large-area global noise reduction. These systems leverage multiple secondary sources and microphones to effectively attenuate noise \cite{iwai2019multichannel,lorente2014gpu}. The conventional centralized strategy requires the processor to not only generate control signals but also update control filters by collecting all inputs, which entails a significant amount of computational effort \cite{Elliott1987MEANC,Douglas1999FXLMS,shi2023computation}.
On the other hand, decentralized control distributes a heavy computational burden among multiple independent controllers. Each controller individually executes a single-channel ANC algorithm \cite{Zhang2019Decentralized,george2012particle}. Nevertheless, disregarding the mutual acoustic crosstalk effect elevates the risk of instability in decentralized systems \cite{Elliott1994InteractionMANC}. Some studies have demonstrated that restricting the maximum noise amplification of independent ANC nodes can enhance overall stability \cite{Zhang2013Decentralized}. However, this method often results in a decline in noise reduction performance. To overcome these limitations, recent research efforts have focused on shaping the matrix eigenvalues of secondary paths for each frequency bin \cite{Pradhan2023DeMCANC,Pradhan2020DeMCANC,Zhang2019Decentralized}. Although these techniques improve performance, they involve complex computational operations and have mainly been proposed for two-channel systems. As an alternative, distributed strategies \cite{Ferrer2015DistributedANC,Chu2020DiffusionANC,Chen2022DistributedANC,Li2023AugmentedDiffusion} aim to address the drawbacks of decentralized strategies by enabling information exchange between ANC nodes. Nevertheless, the introduced communication overhead can have a negative impact on system stability \cite{Ji2023Distributed}.

To address the limitations of distributed ANC without internode communication, i.e. non-collaborative distributed ANC systems, we propose an advanced algorithm that optimizes MCANC performance while minimizing system cost. The core innovation lies in a weight-constrained filtered reference least mean square (WCFxLMS) algorithm, designed to mitigate instability risks caused by inter-node acoustic crosstalk. By dynamically adjusting adaptive weight constraints based on real-time noise reduction performance metrics, each ANC node exhibits self-boosting behavior, ensuring sustained noise attenuation. Compared to conventional decentralized approaches \cite{Zhang2019Decentralized,Pradhan2020DeMCANC,Pradhan2023DeMCANC}, this method demonstrates superior computational efficiency and robustness, supporting different multichannel configurations including the scenarios involving both single and multiple reference microphones.

The remainder of this letter is structured as follows: Section~\ref{sec:method} begins with a concise overview of the conventional algorithms employed for non-collaborative distributed MCANC systems. Subsequently, a detailed exposition of the proposed WCFxLMS algorithm with self-boosted strategy is presented. Section~\ref{sec:sim} evaluates the algorithm's performance through comprehensive simulation studies. Finally, Section~\ref{sec:conclusion} concludes with key findings and contributions while outlining potential future research directions.

\section{Methodology}\label{sec:method}

\subsection{Non-collaborative distributed MCANC}
\begin{figure}[!t]
    \centering
    \includegraphics[width = 0.85\columnwidth]{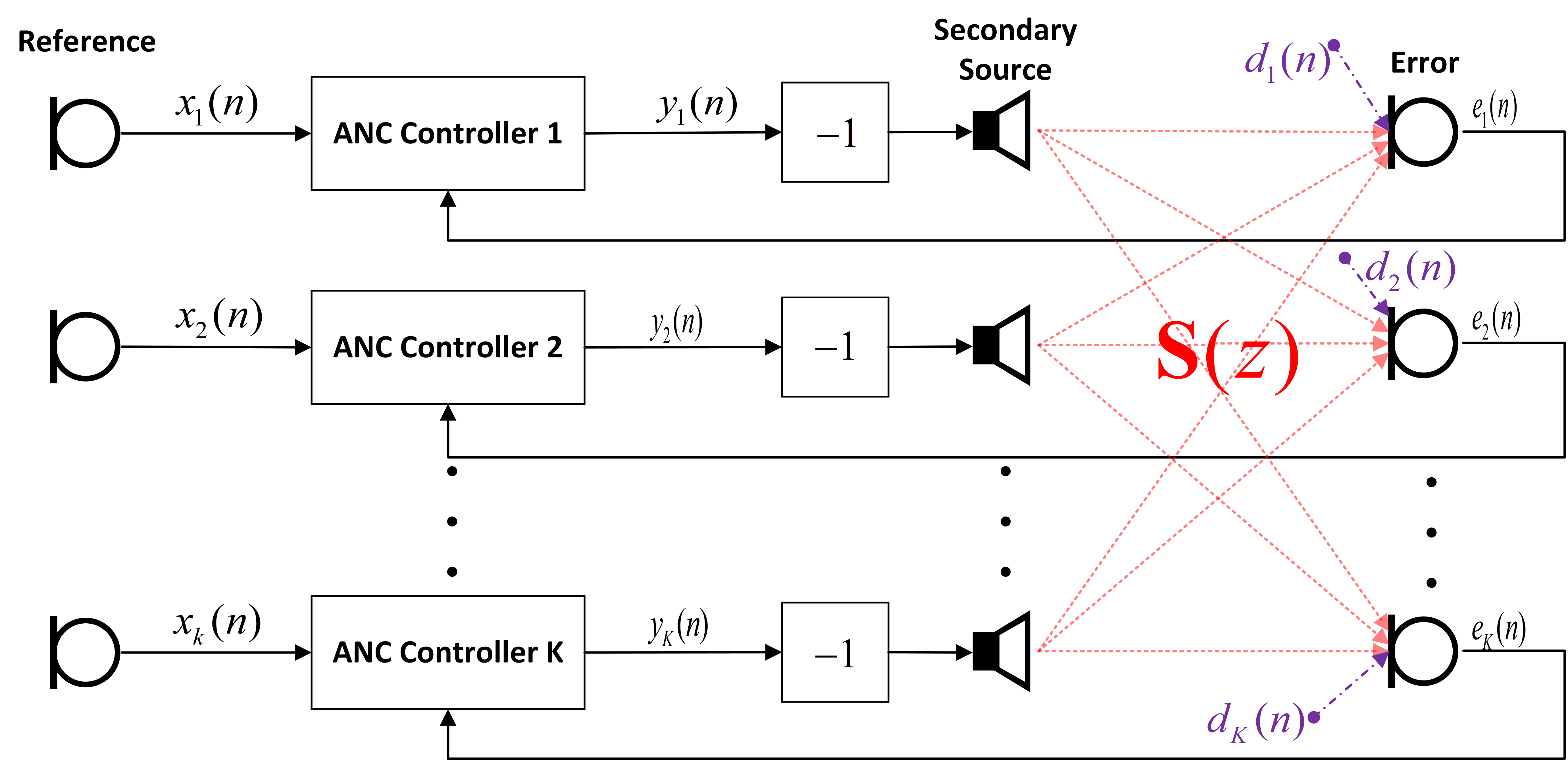}
    \caption{The schematic diagram of MCANC under non-collaborative network, where $\mathbf{S}(z)$ represents the secondary path between the ANC nodes.}
    \label{fig1:DeMCANC}
\end{figure}
Figure \ref{fig1:DeMCANC} illustrates an MCANC system with $K$ independent nodes, each equipped with one reference microphone, one error microphone, and one secondary source.
The control signal of the $k$th node is obtained from
\begin{equation}\label{eq1:controlsignal}
     y_k(n) = \mathbf{w}_k^\mathrm{T}(n)\mathbf{x}_k(n), \quad k = 1,2,...,K,
\end{equation}
where $\mathbf{x}_k(n)=[x_k(n) \, x_k(n-1) \, \cdots \, x_k(n-N+1)]^\mathrm{T}$ denotes the signal vector piked up by the $k$th reference microphone, $\mathbf{w}_k(n)=[w_{k,1}(n) \, w_{k,2}(n) \, \cdots \, w_{k,N}(n)]^\mathrm{T}$ stands for the control filter with $N$ taps, and $n$ is the time index. Meanwhile, the error signal captured by the $k$th error microphone can be expressed as
\begin{equation}\label{eq2:errorsignal}
     e_k(n) = d_k(n) - y_k(n)*s_{kk}(n)-\sum_{m=1,\neq k}^{K}y_m(n)*s_{km}(n),
\end{equation}
in which $*$ denotes the linear convolution, $d_k(n)$ represents the disturbance signal, ${s}_{kk}(n)$ is the impulse response of the inter-node secondary path, and ${s}_{km}(n)$ refers to the impulse response of cross-node secondary paths from the $m$th secondary source to the $k$th error sensor. 

Since each node works independently, its cost function should be
\begin{equation}\label{eq3:Decostfunction}
     J_k = \mathbb{E}[{e}^2_{k}(n)],
\end{equation}
where $\mathbb{E}[\cdot]$ represents the expectation operation. On the basis of the gradient descent method, its negative instantaneous gradient is used to update the control filter:
\begin{equation}\label{eq4:Deupdate}
    \mathbf{w}_k(n+1) = \mathbf{w}_k(n) + \mu \mathbf{x}_{kk}'(n)e_k(n),
\end{equation}
where $\mathbf{x}_{kk}'(n)$ denotes the filtered reference signal vector as
\begin{equation}\label{eq5:filteredx}
    \mathbf{x}_{kk}'(n) = \hat{s}_{kk}(n)*\mathbf{x}_k(n).
\end{equation}
Here, $\hat{s}_{kk}(n)$ represents the estimate of the secondary path $s_{kk}(n)$ with $L$ taps. 
Although this approach saves many computations compared to the conventional MCANC algorithm, it still suffers the instability caused by the cross-talk effect, which can be regarded as an interference, whose expression can be obtained from\eqref{eq2:errorsignal} as
\begin{equation}\label{eq6:interference}
    \gamma_k(n) = \sum_{m=1,\neq k}^{K}y_m(n)*s_{km}(n).
\end{equation}
This interference will introduce a multipath effect that makes the local control filter update unstable or even diverge.

\subsection{Weight-constrained FxLMS (WCFxLMS)}

To constrain the control filter to remain within the vicinity of a specific value, we define the cost function for the $k$th node as follows:
\begin{equation}\label{eq7:CDecostfunction}
     J_k = \mathbb{E}[{e}^2_{k}(n)] + \alpha||\widetilde{\mathbf{w}}_k-\mathbf{w}_k(n)||^2,
\end{equation}
where $\alpha$ and $||\cdot||$ denote the 2-norms operator and penalty factor, respectively, and  $\widetilde{\mathbf{w}}_k$ stands for the center value.

By taking the partial differential of \eqref{eq7:CDecostfunction} with respect to ${\mathbf{w}}_k(n)$, the instantaneous gradient is derived as 
\begin{equation}\label{eq8:gradient}
     \boldsymbol{\nabla}_k = -2\mathbf{x}_{kk}'(n)e_k(n) - 2\alpha(\widetilde{\mathbf{w}}_k-\mathbf{w}_k(n)).
\end{equation}
We then follow the updating rule by using the negative of \eqref{eq8:gradient} to iteratively update the control filter as follows:
\begin{equation}\label{eq9:WCFxLMS}
    \mathbf{w}_k(n+1) = \mathbf{w}_k(n) + \mu \mathbf{x}_{kk}'(n)e_k(n) + \mu \alpha(\widetilde{\mathbf{w}}_k-\mathbf{w}_k(n)).
\end{equation}

Equation~\eqref{eq9:WCFxLMS} is referred to as the WCFxLMS algorithm, as its first two terms coincide with the conventional FxLMS algorithm, while the last penalty term limits the deviation of the adaptive filter from a good point, ensuring system stability. Furthermore, the constraint center point $\widetilde{\mathbf{w}}_k$ can be determined either by pre-training the broadband control filters in advance or by employing artificial intelligence (AI)-based approaches \cite{Luo2023GFANC,Luo2024ImplemSFANC}. As a result, the center point $\widetilde{\mathbf{w}}_k$ encodes prior information about a safe, high‑performing filter. Then the WCFxLMS algorithm trades off between exploiting that prior and exploring adjustments to better track changing noise conditions, while preventing the divergence.

\subsection{Self-boosted strategy}

\begin{figure}[!t]
    \centering
    \includegraphics[width = 0.8\columnwidth]{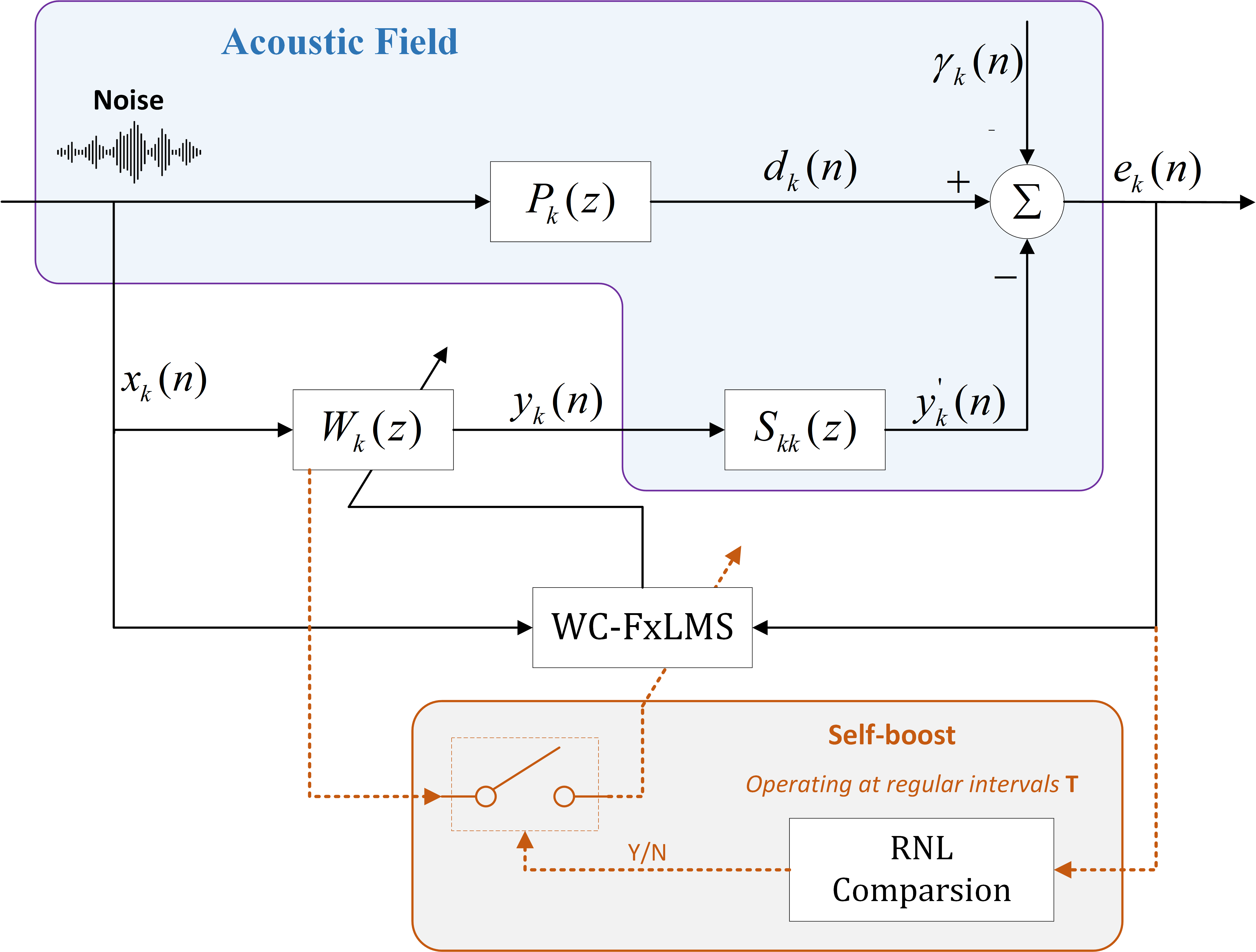}
    \caption{The schematic diagram of the proposed WCFxLMS with self-boosted strategy for the $k$th ANC node, where $P_k(z)$ and $S_{kk}(z)$ represent the primary path and self secondary path, respectively.}
    \label{fig2:STWCFxLMS}
\end{figure}

The WCFxLMS algorithm prevents divergence in decentralized strategies by constraining the range of control filter updates. However, this limitation can occasionally hinder individual ANC nodes from achieving optimal noise reduction, as restricting the update range inevitably compromises performance to some degree. To overcome this issue, the center point $\widetilde{\mathbf{w}}_k$ can be adjusted, prompting the algorithm to re-explore potentially more effective solutions. Thus, a self-boosted strategy is proposed, periodically updating the center point $\widetilde{\mathbf{w}}_k$ to new positions, thereby enabling the system to achieve enhanced noise reduction performance.

When using the WCFxLMS algorithm for noise suppression, the center point $\widetilde{\mathbf{w}}_k$ can be periodically updated with the current control filter value $\mathbf{w}_k(n)$. However, to maintain system stability, the residual noise level (RNL) of each ANC node should be evaluated before replacing the constrained center point $\widetilde{\mathbf{w}}_k$ with the latest control filter $\mathbf{w}_k(n)$. Typically, the RNL $\eta_k(n)$ is derived from the residual signal $e_k(n)$, defined as follows:
\begin{equation}\label{eq10:nrl}
    \eta_k(n) = 10\log_{10}\left[{e^2_k(n)}\right].
\end{equation}

Due to the slow convergence of the LMS algorithm, the system may require considerable time to find an improved solution. The proposed algorithm may introduce additional noise by frequently updating the center point $\widetilde{\mathbf{w}}_k$. Hence, an updating period $T$ is defined, during which the average residual noise level (RNL) is computed to be
\begin{equation}\label{eq11:anrl}
    \bar{\eta}_k = \frac{1}{Tf}\sum_{t=n-Tf}^{n} \eta_k(t),
\end{equation}
where $f$ denote the sampling frequency in Hz. Additionally, we also define the minimum average RNL as $\bar{\eta}_{k,min}$. When the current average noise reduction outperforms the previously recorded best result, i.e., 
$ \bar{\eta}_k < \bar{\eta}_{k,min}$, the current control filter is assigned as the new center point $\widetilde{\mathbf{w}}_k$, and the minimum average RNL $\bar{\eta}_{k,min}$ is updated accordingly.

Combined with this self-boosted strategy, the proposed self-boosted WCFxLMS (SB-WCFxLMS) algorithm can be illustrated in Fig.~\ref{fig2:STWCFxLMS}. For each ANC node, the WCFxLMS algorithm is implemented to prevent system divergence caused by crosstalk from other ANC nodes. The average noise reduction level is computed at regular intervals T. If the current noise reduction surpasses the previously recorded best value, the current control filter is set as the new center point filter in the WCFxLMS algorithm, and the optimal noise reduction level is updated accordingly. By continuously refining these constraints, the system progressively converges toward better control filters, thus achieving enhanced noise reduction performance. The pseudo-code of the proposed SB-WCFxLMS algorithm is detailed in Table~\ref{tab:algorithm}.

Furthermore, the proposed algorithm requires less computational effort compared to the centralized approach and reduces the communication burden inherent in distributed MCANC schemes. Additionally, it enhances both the stability and noise reduction performance of multiple ANC nodes operating within non-collaborative distributed networks.

\begin{table}[!t]
    \centering
    \caption{Pseudo-code of SB-WCFxLMS algorithm}\label{tab:algorithm}
    \begin{tabular}{l}
    \hline
    \textbf{Algorithm:} SB-WCFxLMS algorithm for the $k$th ANC node\\
    \hline
    \textbf{Initialization:} Obtain $\hat{\mathbf{s}}_{kk}(n)$ and $\widetilde{\mathbf{w}}_k$.\\
    \textbf{Input:} The reference signal $x_k(n)$; The error signal $e_k(n)$;\\
    \textbf{Output:} The control signal $y_k(n)$.\\
    \textbf{While} True \textbf{do}\\
     ~~~$y_k(n) \leftarrow \mathbf{x}_k^\mathrm{T}(n)\mathbf{w}_k(n)$\\
     /*WCFxLMS */\\
     ~~~$\mathbf{x'}_{kk}(n) \leftarrow \hat{s}_{kk}(n) * \mathbf{x}(n)$ \\
     ~~~$\mathbf{\hat{w}}(n+1) \leftarrow  \mathbf{w}_k(n) + \mu \mathbf{x}_{kk}'(n)e_k(n) + \mu \alpha(\widetilde{\mathbf{w}}_k-\mathbf{w}_k(n))$\\
     /*Self-boosted*/\\
     ~~~$\eta_k(n) \leftarrow 10\log_{10}\left[{e^2_k(n)}\right]$ \\
     ~~~\textbf{if} $\mod(n,Tf) == 0$ ~~~~~~~~~~~~$\triangleright$ At intervals of time $T$ \\
     ~~~~~ $\bar{\eta}_k \leftarrow \frac{1}{Tf}\sum_{t=n-Tf}^{n} \eta_k(t)$\\
     ~~~~~ \textbf{if} $ \bar{\eta}_k < \bar{\eta}_{k,min}$\\
     ~~~~~~~ $\widetilde{\mathbf{w}}_k \leftarrow \mathbf{w}_k(n) $\\
     ~~~~~~~ $\bar{\eta}_{k,min} \leftarrow \bar{\eta}_k $\\
     ~~~~~ \textbf{end if}\\
     ~~~\textbf{end if}\\
     \textbf{end while}\\
    \hline
    \end{tabular}
\end{table}

\section{Numerical Simulations}\label{sec:sim}
In this section, we validate the performance of our proposed SB-WCFxLMS algorithm in an MCANC system consisting of $6$ ANC nodes. The primary and secondary paths are measured within a noise chamber equipped with an ANC window, and the system configuration is the same as \cite{ji2025mixed}. The tap lengths for the control filters and secondary path modeling are set to $512$ and $256$, respectively, the control filters are all initialized to zero, and the sampling frequency is $16,000$Hz. The initial center point filter $\widetilde{\mathbf{w}}_k$ is initialized to zero, making the initial state equivalent to the Leaky FxLMS algorithm \cite{Kuo1999ANC}. The residual noise level (RNL) is computed at one-second intervals. For the single-reference scenario, the proposed SB-WCFxLMS algorithm is compared with the conventional centralized MCANC \cite{Elliott1987MEANC}, decentralized MCANC (DeMCANC) \cite{Kuo1999ANC}, Augmented diffusion FxLMS (ADFxLMS) \cite{Li2023AugmentedDiffusion}, and Leaky FxLMS algorithms. For the multi-reference scenario, comparisons are made against the collocated MCANC \cite{shi2016open}, multi-reference DeMCANC (MRDeMCANC), and Leaky FxLMS algorithms. To quantify noise reduction performance, the average normalized squared error (ANSE) across all ANC nodes is defined as
\begin{equation}\label{eq12:anse}
    \text{ANSE} = \frac{1}{K}\sum_{k=1}^{K} 10\log_{10}\bigg\{\frac{\mathbb{E}[e^2_k(n)]}{\mathbb{E}[d^2_k(n)]}\bigg\}.
\end{equation}

\subsection{Noise reduction performance on multi-tonal noise}
\begin{figure}[!t]
    \centering
    \includegraphics[width = 0.85\columnwidth]{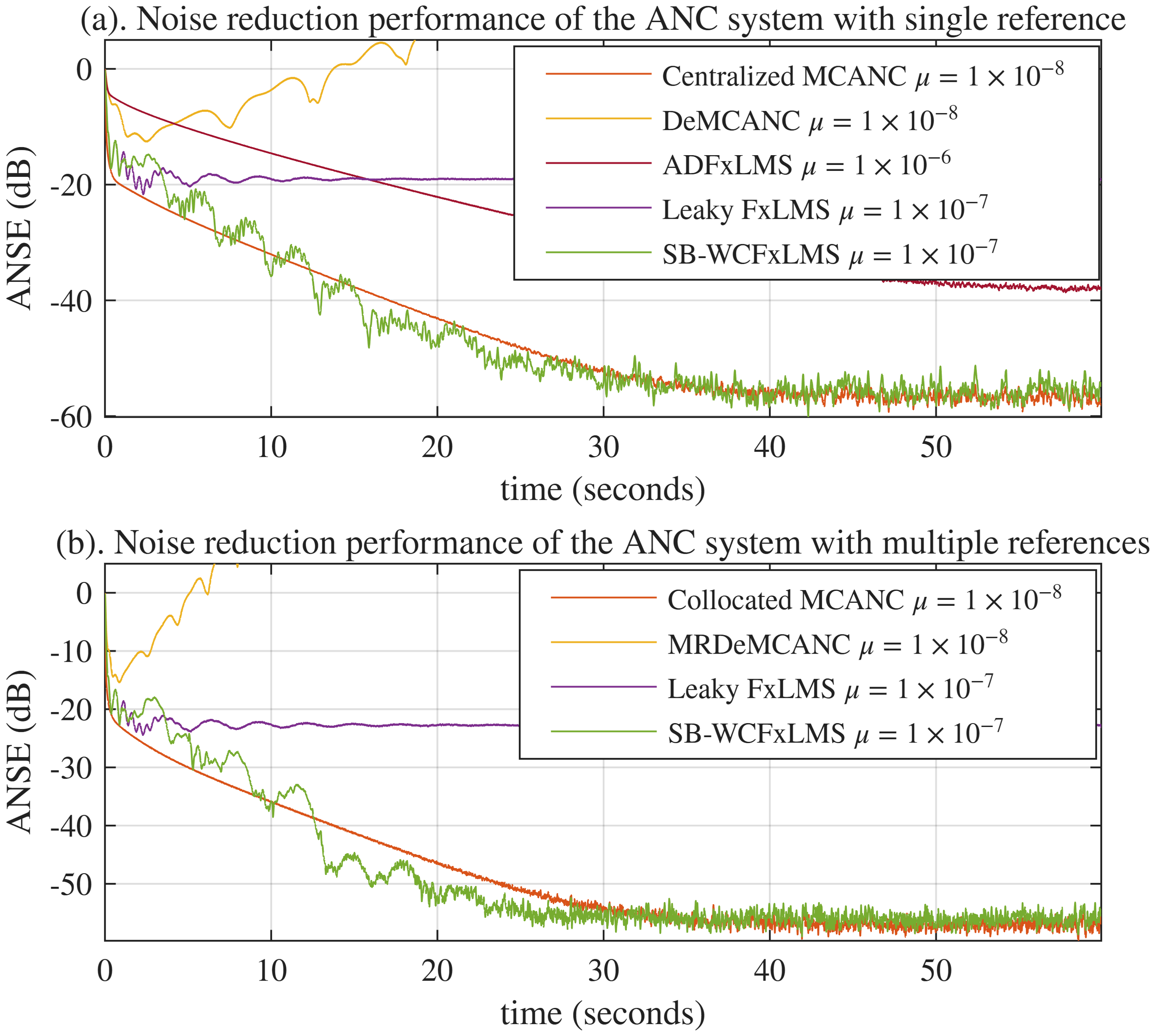}
    \caption{Noise reduction performance for multi-tonal noise: (a) ANSE comparison of various ANC systems with a single shared reference signal across all ANC nodes; (b) ANSE comparison of various ANC systems, where each ANC node uses an individual reference signal.}
    \label{fig4:case2}
\end{figure}

To validate the proposed algorithm, the primary noise is chosen as a multi-tonal signal composed of frequencies at $300$, $400$, $500$, $600$, and $700$ Hz. The penalty factor $\alpha$ is chosen as $1000, 1000, 2000, 2000, 1000, 1000$ for each node. As illustrated in Fig.~\ref{fig4:case2}(a), the proposed SB-WCFxLMS algorithm achieves noise reduction performance nearly equivalent to that of the centralized algorithm. In contrast, the decentralized strategy fails to converge due to crosstalk effects among the ANC nodes. The Leaky FxLMS algorithm prevents divergence but exhibits limited noise reduction performance because of constraints on the filter update range. The ADFxLMS algorithm averages information received by the nodes, which slightly compromises noise reduction. 

For the multi-reference scenario shown in Fig.~\ref{fig4:case2}(b), the proposed SB-WCFxLMS again demonstrates performance comparable to the collocated MCANC method. Although the proposed SB-WCFxLMS algorithm exhibits fluctuations during the convergence phase due to its self-boosted mechanism, it ultimately reaches the same steady-state performance as the baseline algorithm. Overall, these results confirm that the proposed algorithm provides satisfactory performance for multi-tonal noise cancellation.


\subsection{Noise reduction performance on real-recorded noises}

\begin{figure}[!t]
    \centering
    \includegraphics[width = 0.85\columnwidth]{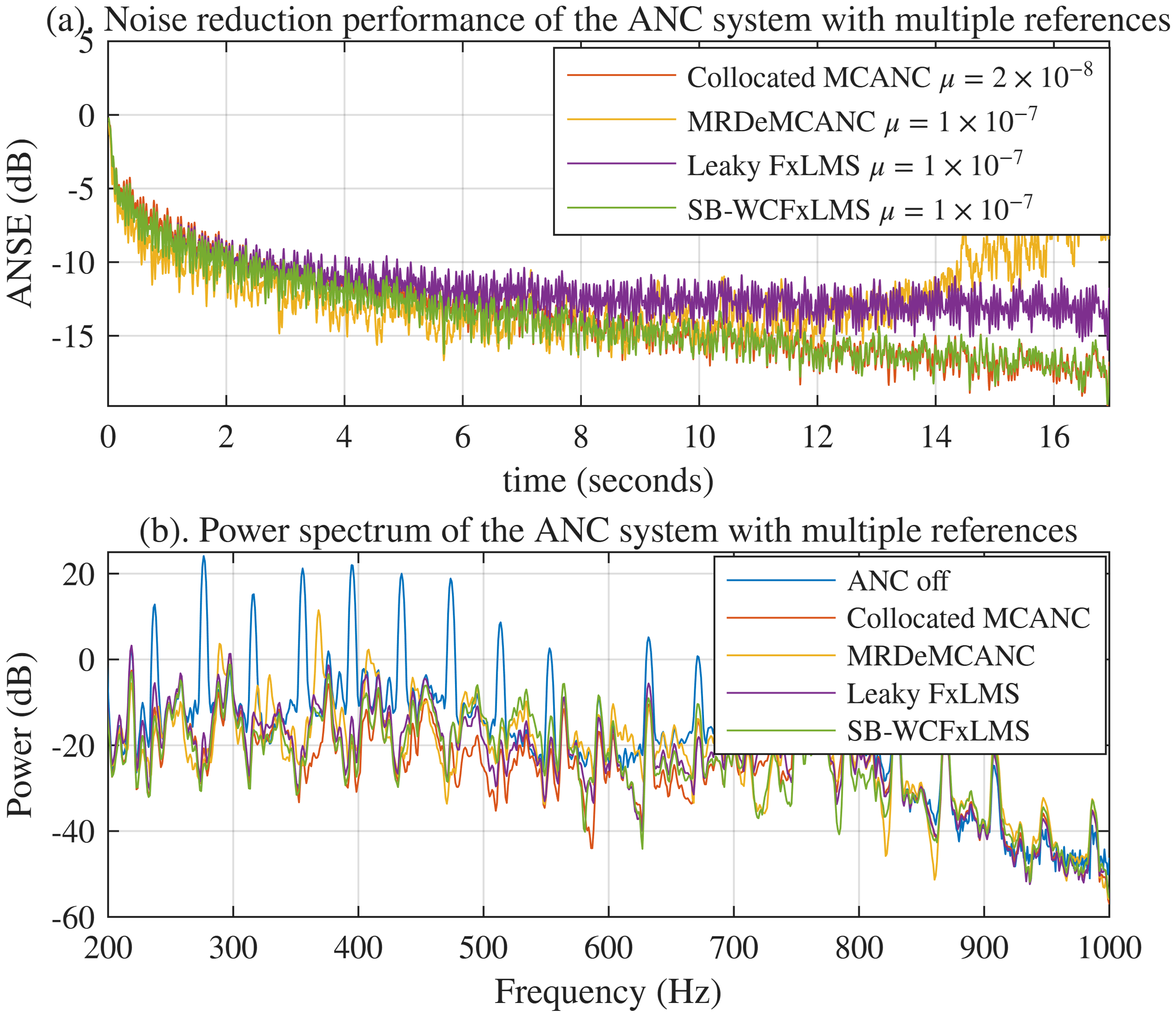}
    \caption{Noise reduction performance for real recorded compressor noise: (a) ANSE comparison of various MCANC systems, where each ANC node uses an individual reference signal; (b) Power spectrum comparison of various MCANC algorithms}
    \label{fig5:case3}
\end{figure}

\begin{figure}[!t]
    \centering
    \includegraphics[width = 0.85\columnwidth]{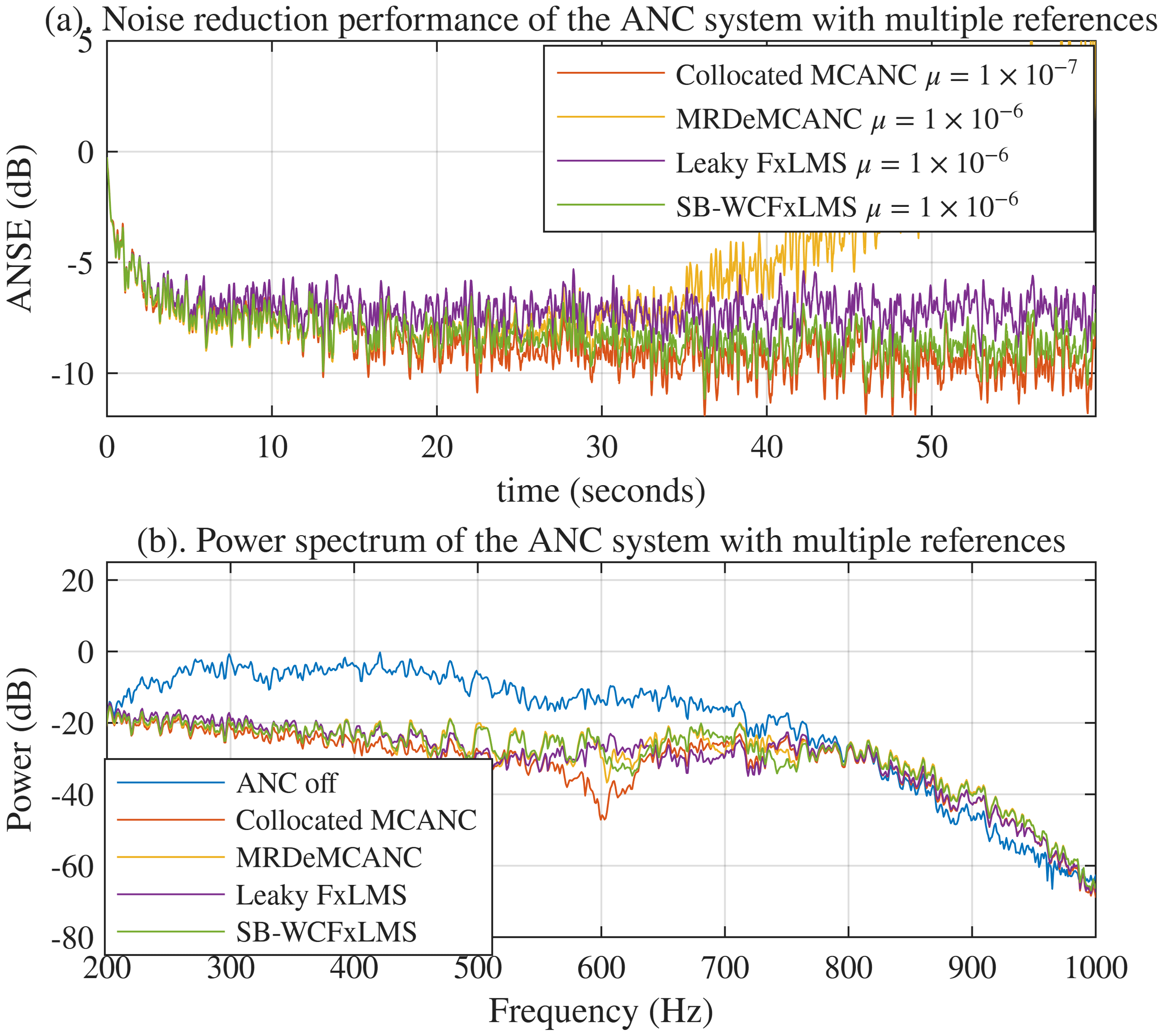}
    \caption{Noise reduction performance for broadband noise: (a) ANSE comparison of various MCANC systems, where each ANC node uses an individual reference signal; (b) Power spectrum comparison of various MCANC algorithms}
    \label{fig5:case4}
\end{figure}

In the first simulation, a real recorded compressor noise was selected as the primary noise source. The $\alpha$ is chosen as $100$ for all the nodes. As depicted in Fig.~\ref{fig5:case3}, the decentralized algorithms gradually diverge, and the leaky strategy fails to achieve satisfactory noise reduction performance. In contrast, the proposed SB-WCFxLMS algorithm achieves performance comparable to the traditional centralized approach under real environmental conditions. In terms of the noise reduction performance of the broadband noise from 200 to 600Hz shown in Fig.~\ref{fig5:case4}, the proposed SB-WCFxLMS is slightly worse than the baseline since prior information from other nodes may be absent where the penalty factor is set to $10$. However, it still outperforms the other algorithms, and the power spectrum shows that the main frequency components are attenuated.

\section{Conclusion}\label{sec:conclusion}

This paper introduces a self-boosted weight-constrained FxLMS (SB-WCFxLMS) algorithm designed for distributed MCANC systems. The proposed approach eliminates the need for inter-node communication and effectively handles multiple reference inputs. By employing weight constraints, the algorithm mitigates the 'crosstalk' effects between outputs of different nodes and enhances the stability of the distributed system. Moreover, a self-boosted strategy continuously optimizes the constraint center point, enabling the distributed MCANC system to achieve noise reduction performance comparable to conventional methods. Numerical simulations conducted in a realistic environment verify the effectiveness of the proposed SB-WCFxLMS algorithm in both single-reference and multi-reference scenarios, demonstrating substantial noise reduction without the assistance of the internode communication. Therefore, the proposed approach holds significant potential for practical application in distributed MCANC systems, enhancing both stability and noise reduction performance while simultaneously reducing communication overhead.

\ifCLASSOPTIONcaptionsoff
  \newpage
\fi

\newpage

\bibliographystyle{IEEEtran}
\bibliography{./bibtex/bib/IEEEabrv,./bibtex/bib/ref}

\end{document}